\documentclass[12pt,twoside]{article}   
\usepackage[super,sort,comma]{natbib}

\usepackage{fancyhdr}		

\usepackage[section]{placeins}   %

\usepackage{graphicx}
\usepackage{amsmath,amssymb,amsfonts}
\usepackage{algorithmicx}
\usepackage{textcomp}
\usepackage{algorithm}
\usepackage{algpseudocode}

\DeclareMathOperator*{\solve}{solve}
\DeclareMathOperator{\sh}{shrink}
\DeclareMathOperator*{\argmin}{arg\,min}

\makeatletter \renewcommand\@biblabel[1]{$^{#1}$} \makeatother
\newcommand{\note}[1]{\mbox{}\\ \noindent \rule{16cm}{0.5mm} \\
{\em #1} \\ \noindent \rule{16cm}{0.5mm}
\typeout{    }
\typeout{***********note active on this page *************************}
\typeout{Note: #1  }
\typeout{****************************************end Note}
}

\renewcommand{\note}[1]{}

\newcommand{\cen}[1]{\begin{center} #1 \end{center}}

\usepackage[mathlines]{lineno}

\usepackage{hyperref}
\hypersetup{ colorlinks,
    citecolor=blue,
    filecolor=blue,
    linkcolor=blue,
    urlcolor=blue
}

\usepackage{xcolor}

\definecolor{gray}{rgb}{0.6,0.6,0.6}
\definecolor{red}{rgb}{0.85,0,0}
\definecolor{green}{rgb}{0,0.85,0}
\definecolor{blue}{rgb}{0,0,0.85}
\definecolor{beige}{rgb}{0.92,0.87,0.78}
\usepackage[all]{hypcap}    

\begin{document}

\cen{\sf {\Large {\bfseries A signal detection model for quantifying over-regularization in non-linear image reconstruction } \\  
\vspace*{5mm}
Emil Y. Sidky$^1$, John Paul Phillips$^1$, Weimin Zhou$^2$, Greg Ongie$^3$, Juan P. Cruz-Bastida$^1$, Ingrid S. Reiser$^1$, Mark A. Anastasio$^2$, and Xiaochuan Pan$^1$
 } \\
$^1$Department of Radiology, The University of Chicago, 5841 S. Maryland Ave., Chicago, IL 60637, USA\\
$^2$Department of Bioengineering, University of Illinois at Urbana-Champaign, 1406 W. Green St., Urbana, IL 61801, USA\\
$^3$Department of Mathematical and Statistical Sciences, Marquette University, 1313 W. Wisconsin Ave., Milwaukee, WI 53233, USA
\vspace{3mm}\\
Version typeset \today\\
}

\pagenumbering{roman}
\setcounter{page}{1}
\pagestyle{plain}
Author to whom correspondence should be addressed. email: sidky@uchicago.edu\\

\begin{abstract}
\noindent {\bf Purpose:} Many useful image quality metrics for evaluating linear image reconstruction techniques
do not apply to or are difficult to interpret for non-linear image reconstruction.  
The vast majority of metrics employed for evaluating non-linear image reconstruction are based on
some form of global image fidelity, such as image root mean square error (RMSE). Use of such metrics
can lead to over-regularization in the sense that they can favor removal of subtle details in
the image. To address this short-coming, we develop an image quality metric based
on signal detection that serves as a surrogate to the qualitative loss of fine image details.\\
{\bf Methods:} The metric is demonstrated in the context of a breast CT simulation, where different equal-dose
configurations are considered. The configurations differ in the number of projections acquired.
Image reconstruction is performed with a non-linear algorithm based on total variation constrained
least-squares (TV-LSQ). The resulting images are studied as a function of three parameters: number of views acquired,
total variation constraint value, and number of iterations. The images are evaluated visually, with image
RMSE, and with the proposed signal-detection based metric. The latter uses a small signal, and computes
detectability in the sinogram and in the reconstructed image. Loss of signal detectability through the image
reconstruction process is taken as a quantitative measure of loss of fine details in the image.\\
{\bf Results:} Loss of signal detectability is seen to correlate well with the blocky or patchy appearance due to 
over-regularization with TV-LSQ, and this trend runs counter to the image RMSE metric, which tends to favor
the over-regularized images.\\
{\bf Conclusions:} The proposed signal detection based metric provides an image quality assessment that is complimentary
to that of image RMSE. Using the two metrics in concert may yield a useful prescription for determining CT
algorithm and configuration parameters when non-linear image reconstruction is used.\\
\end{abstract}
\note{This is a sample note.}

\newpage     

\tableofcontents

\newpage

\setlength{\baselineskip}{0.7cm}      

\pagenumbering{arabic}
\setcounter{page}{1}
\pagestyle{fancy}

\section{Introduction}
\label{sec:introduction}

The effort in developing non-linear image reconstruction algorithms for X-ray
computed tomography (CT) has been steadily
increasing over the past couple of decades. The non-linearity arises from incorporation
of some forms of prior information in the reconstruction process or some forms of physics modeling.
For example, edge-preserving
regularization and spectral response modeling both yield an image reconstruction algorithm
that yields images that depend non-linearily on the CT data \cite{elbakri2002statistical,mccollough2009strategies}. Exploitation
of sparsity or transform sparsity also involves non-linear image reconstruction \cite{sidky2008image,chen2008prior,ritschl2011improved,batenburg2011dart}.
Most recently,
deep-learning based data processing is being investigated for generating tomographic images directly
from CT projection data using convolutional neural networks (CNNs) \cite{gupta2018cnn,adler2018learned}.
Such CNNs also process the tomographic
data in a non-linear fashion.

While non-linear image reconstruction may allow for accurate image reconstruction
in CT systems involving low-dose illumination or sparse sampling, the resulting image characteristics
can depend strongly on the scanned object. This object dependence presents a difficult challenge
for developing meaningful image quality metrics needed to guide algorithm parameter selection in a
non-subjective fashion. As a result, much work on non-linear image reconstruction techniques present
images resulting from algorithms where the parameters are tuned by eye. Such an approach may be fine
for an initial introduction of a new image reconstruction algorithm or if the CT system/reconstruction
parameter space is limited enough where it is feasible to tune by eye. The tune by eye method, however,
blunts the impact of advanced image reconstruction because such image reconstruction techniques themselves
involve numerous parameters and they
aim to broaden the scope of possible CT system configurations -- enlarging the parameter space of CT hardware.
Attempting to perform comparisons between different non-linear image reconstruction algorithms only complicates
the matter further. With a large parameter space, the tune-by-eye method becomes impractical.

Avoiding the subjective tune-by-eye method, many researchers in advanced CT image reconstruction turn to
one of three image fidelity metrics in their simulations: root-mean-square-error (RMSE), peak signal-to-noise ratio (PSNR),
or structural similarity index (SSIM). These metrics are useful, in a simulation setting, because they present a measure
of how close a reconstructed image is to a ground truth image. This information in turn is useful for investigating the
underlying inverse problem. When considering clinical imaging tasks that rely on viewing subtle image features, optimizing
system and reconstruction parameters on these global image fidelity metrics
can lead to significantly over-regularized images.

One problem is that 
these image fidelity metrics do not provide a sense of image resolution, noise
level, or noise quality. PSNR, from its name, would seem to provide information on the image noise level, but what is
called ``noise'' in PSNR is actually the difference between the reconstructed and truth images, and this difference includes
both image noise and deterministic artifacts from either unmodeled non-stochastic physics or insufficient sampling.
For non-linear image reconstruction algorithms, concepts such as the point-spread function and the noise power spectrum
do not have a simple and direct interpretation as they do for linear systems theory. For example, in
non-linear image reconstruction,
the resulting image cannot be interpreted as a convolution of a reconstructed point-like object and the underlying
true object function. As a result, they are used rarely
in the evaluation of non-linear image reconstruction. 

In order to prevent over-smoothing by optimizing non-linear image reconstruction solely on image RMSE, an image quality metric
is needed that is sensitive to subtle features in the image and that is easy to interpret. To develop such a metric, we turn
to signal detection theory, and investigate the use of the ideal observer for a simple signal-known-exactly/background-known-exactly
detection task \cite{barrett2004foundations}. Signal detection theory has been investigated in the context of evaluation of image
reconstruction algorithms \cite{abbey1996observer,abbey2001human,wunderlich2008image,das2010penalized,sanchez2013comparison,gang2014task,sanchez2014task,xu2015task,rose2017investigating}.
For the present work, the signal is chosen to be a point-like object and its amplitude is set so that it is at the limit of detectability in
the CT data space. It is known, that image reconstruction or any other image processing operations cannot increase signal detectability
(see pages 829-30 in Barrett and Myers\cite{barrett2004foundations}),
but it is possible that image reconstruction can reduce signal detectability. Quantifying this loss of detectability is precisely what
we would like to use as a measure of over-regularization. Having such a measure would allow optimization of image RMSE under the condition
that signal detection is constrained to be at or above a desired set level and thus prevent over-regularization.

The setting for developing this metric, here, is a dedicated breast CT simulation where image reconstruction is performed
by total-variation (TV), least-squares optimization (TV-LSQ). The TV-LSQ algorithm is non-linear and it allows for accurate
image reconstruction from sparse-view CT data under ideal noiseless conditions. When TV-LSQ is employed for noisy, realistic
data it is often reported that the images are patchy or blocky, and one solution to avoid this subject quality is
to generalize the TV-norm  \cite{liu2013total,niu2014sparse}. For the present work, however, we argue that the patchiness
resulting from use of TV regularization can also be a side effect of over-regularization due to parameter optimization using
image RMSE. Using the proposed signal detectability metric can help to disallow parameter settings that cause over-regularization
and, specifically, the patchy appearance from over-regularization with the TV-norm.

We point out that the patchy appearance for over-regularization with the TV-norm is a somewhat subjective assessment, and
therefore the claim that the proposed metric characterizes patchiness quantitatively is also subjective and cannot be proven
mathematically. We do attempt to design the simulation so that the subjectivity is limited as much as possible, but in the
end the utility of the proposed metric can only be demonstrated by showing metric correspondences with images and it is left
to the observer to decide whether this correspondence is useful or not.

In Sec. \ref{sec:methods}
we present the parameters of the breast CT simulation, the details of the TV-LSQ
algorithm, and the channelized Hotelling observer (CHO) for the signal-known-exactly/background-known-exactly (SKE/BKE) detection task.
The results, presented in Sec. \ref{sec:results},
demonstrate the correspondences between the proposed signal detection metric and reconstructed images for select parameter
settings of the breast CT simulation and TV-LSQ algorithm. The results are discussed in Sec. \ref{sec:discussion}, and
finally, we conclude the paper in Sec. \ref{sec:conclusion}.

\section{Methods}
\label{sec:methods}

\subsection{Breast CT simulation}
For the studies presented here, we consider
a fixed dose simulation, where the number of projections is varied while keeping the total patient exposure constant.
The configuration is 2D circular, fan-beam scanning and is representative of the mid-plane slice of a 3D circular cone-beam scan.
The mean continuous data function, $g$, is modeled as the X-ray transform of the object function
\begin{linenomath}
\begin{equation}
\label{xray}
g(\theta,\xi)=Pf(\theta,\xi) = \int_0^\infty f(r_0(\theta) + t \, \hat{\phi}(\theta,\xi)) dt,
\end{equation}
\end{linenomath}
where $f$ represents the continuous object function; $Pf$ is the continuous X-ray transform of $f$;
 $\theta$ indicate the view angle of the X-ray source;
$\xi$ is the detector bin location on a linear detector; $r_0(\theta)$ indicates the X-ray source position; and the unit
vector $\hat{\phi}$ points from the X-ray source to the detector bin indicated by $\xi$, accordingly $\hat{\phi}$ is a function
of $\theta$ and $\xi$. The data function is sampled at a variable number of views $N_\text{views}$ and 512 detector bins.
The noise level in the measured transmission data is specified by fixing the total number of incident photons to
\begin{linenomath}
\begin{equation*}
N_\text{photons} = 10^{10}.
\end{equation*}
\end{linenomath}
In the simulations we consider varying $N_\text{views}$ between 128 and 512, and for the maximum end of this
range the number of incident photons along each measured ray is approximately 38,000 photons, which is on the low
end of actual breast CT systems \cite{boone2005technique,sanchez2014task}. To model noise due to the detection of finite numbers of quanta,
a Poisson distribution is assumed in the X-ray transmission measurements. Accounting for the logarithm processing
needed to arrive at the line-integration model, Eq. (\ref{xray}), we model the noisy discrete data with a Gaussian
distribution with mean
\begin{linenomath}
\begin{equation}
\label{datamean}
\bar{g_\ell} = g(\theta_\ell,\xi_\ell),
\end{equation}
\end{linenomath}
and variance
\begin{linenomath}
\begin{equation}
\label{datavar}
\text{Var}(g_\ell) =  \left(\frac{N_\text{photons}}{N_\text{views}} \exp(-g(\theta_\ell,\xi_\ell))\right)^{-1},
\end{equation}
\end{linenomath}
where $\ell$ is an index for each of the transmission rays in the projection data.
It is clear from Eq. (\ref{datavar}) that the noise variance decreases with decreasing numbers of views,
and there is a tradeoff between $N_\text{views}$ and signal-to-noise ratio in each projection.

\subsection{TV-LSQ image reconstruction}

In order to formulate the TV-LSQ optimization the continuous data model in Eq. (\ref{xray})
is discretized, taking the form of a large linear system
\begin{linenomath}
\begin{equation*}
g=Xf,
\end{equation*}
\end{linenomath}
where the pixelized 512$\times$512 image is represented by $f$; X-ray projection becomes the matrix $X$;
and the $N_\text{views} \times 512$ data is denoted by $g$. Because we consider $N_\text{views} \le 512$ this linear
system can be under-determined.
The TV-LSQ optimization problem is formulated as
\begin{linenomath}
\begin{equation}
\label{TVLSQ}
f^\star = \argmin_f \frac{1}{2} \| g-Xf\|^2_2 \text{ such that } \|(|Df|_\text{mag})\|_1 \le \gamma,
\end{equation}
\end{linenomath}
where $D$ is the finite differencing approximation to the image gradient; $|\cdot|_\text{mag}$ is the pixelwise
magnitude of the spatial gradient vector $Df$; $\|(|Df|_\text{mag})\|_1$ is the image total variation (TV);
and $\gamma$ is the TV constraint value. When the data $g$ are generated from a test image $f_\text{test}$ with no noise added,
the test image can be recovered with TV-LSQ choosing $\gamma=\|Df_\text{true}\|_1$ for
sparse-view sampling with $N_\text{views} < 512$. The degree of under-sampling permitted depends on the sparsity
in the gradient magnitude image (GMI)  $|D f_\text{test}|_\text{mag}$ \cite{jorgensen2015little}.
This possibility of accurate image reconstruction for sparse-view CT enables the consideration of the CT configurations 
described in the breast CT simulation.

The TV-LSQ optimization problem can be efficiently solved by the Chambolle-Pock primal-dual (CPPD)
algorithm \cite{chambolle2011first,Pock2011,SidkyCP:12}.
For completeness we provide the pseudocode for this algorithm in Appendix \ref{app:cppd}.
We do consider early stopping of the CPPD algorithm and allow the total number
of iterations, $N_\text{iter}$, to vary from 10 to 500. At $N_\text{iter}=500$
the TV-LSQ problem is solved to a high degree of numerical accuracy for all scan configurations considered in this work.

In total, three parameters are varied in the breast CT simulation: $N_\text{views}$, $N_\text{iter}$, and the TV constraint $\gamma$.
Even under this restricted simulation with three parameters specifying the image,
it is difficult to tune-by-eye; not only is the parameter space too large but the image qualities
are difficult to compare. As will be seen, quantitative image fidelity metrics such as image RMSE,
alone, may not provide a reasonable objective means
of image comparison and optimization, particularly when small subtle signals are the features of interest.

\subsection{SKE/BKE signal detection model}

To provide an objective metric that characterizes the preservation of subtle details in the TV-LSQ reconstructed images,
signal detection theory is employed to measure the loss of signal detectability for an ideal observer model.
The design of the detectability metric involves the following steps:
select the signal properties such that it is on the border
of detectability in the sinogram data domain; generate multiple realizations of signal-present and signal-absent sinograms;
perform TV-LSQ reconstruction of all data realizations; divide the resulting image set into training and testing data;
train the signal-present/signal-absent classifier; and finally, measure the image domain detectability with the testing images.
The data model and data signal detection task is set up so that the ideal observer performance can be analytically computed.
In this way, the data domain detectability serves as a precisely known upper bound to the image domain detectability. The loss in detectability,
passing through image reconstruction, provides a quantitative measure that is an indication of loss of fine details in the image
and may reflect the subjective property of image patchiness.

\begin{figure}[!t]
\centerline{\includegraphics[width=0.5\columnwidth]{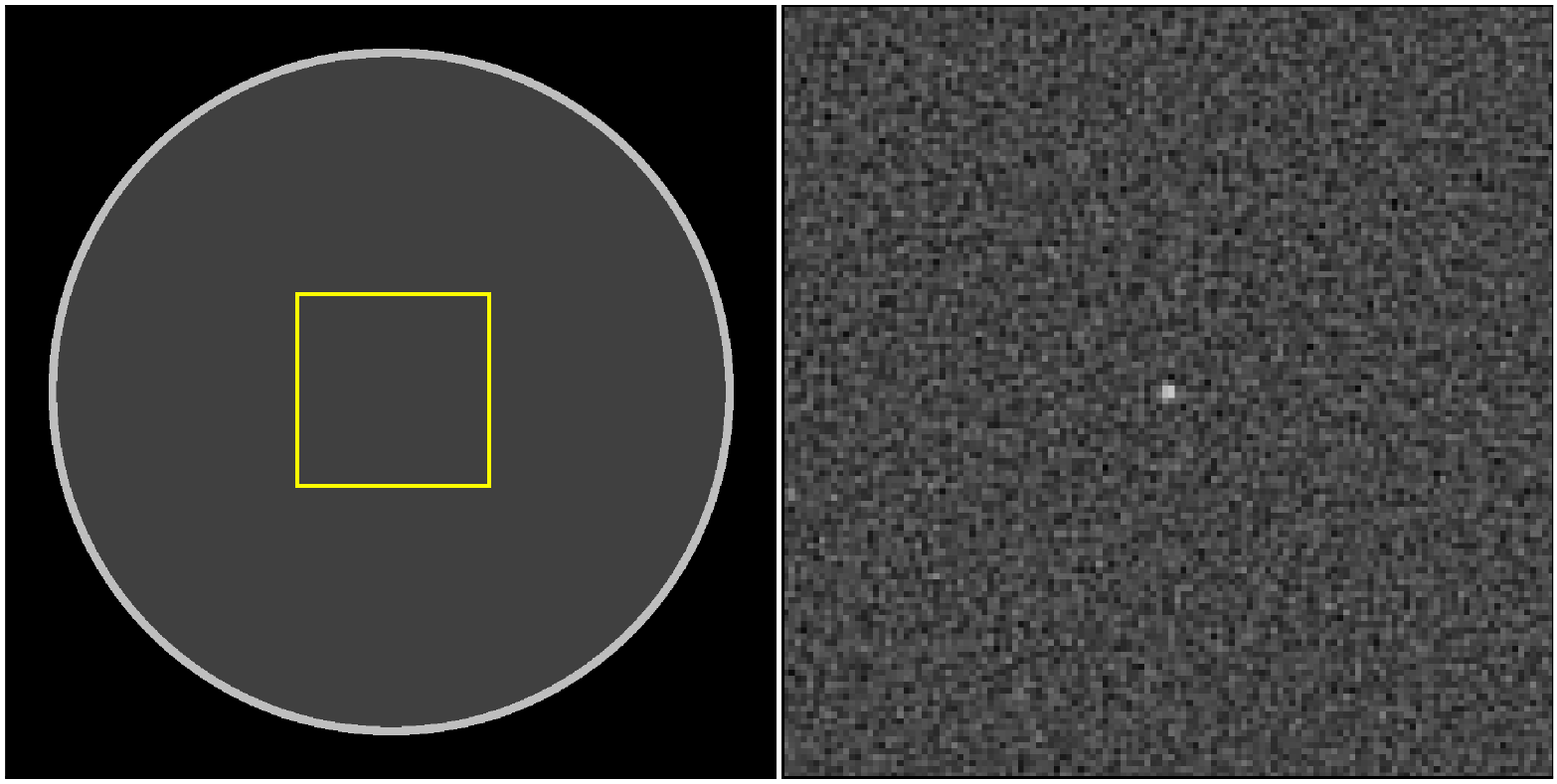}}
\caption{(Left) Background image used for the signal-known-exactly/background-known-exactly detection task.
The gray scale window is [0.174,0.253]  cm$^{-1}$. 
(Right) Central 128$\times$128 ROI of the mean difference of 200 filtered back-projection (FBP) reconstructed noise realizations
from the signal-present and signal-absent sinograms. The size of this central ROI is indicated with the yellow box on the background image.
The gray scale window is [-0.0075,0.02]  cm$^{-1}$. 
}
\label{fig:detection}
\end{figure}

The images in Fig. \ref{fig:detection} illustrate the detection task employed in this work.
The background disk attenuation is representative of fat tissue and is set to 0.194 cm$^{-1}$. The ring at the edge represents
the skin-line with attenuation 0.233 cm$^{-1}$. 
The phantom is defined on a 2048$\times$2048 grid and is 18$\times$18 cm$^{2}$
in physical dimensions. The pixel size is chosen much smaller than the detector resolution so that the phantom
can be regarded as quasi-continuous. Projection of this background image yields the mean background sinogram.
The signal is defined as a Gaussian function with full-width-half-maximum of 100 microns (the reconstructed image
grid uses a pixel size of 350 microns) and amplitude of 0.04 cm$^{-1}$. Projection of the background plus signal
yields the mean signal-present sinogram.
To appreciate the difficulty of the detection task, we also show in Fig. \ref{fig:detection}
the mean difference image of both hypotheses
over 200 noise realizations, reconstructed by FBP for $N_\text{views}=512$.
The reconstruction grid is a 512$\times$512 pixel array.
It is apparent that the speckle noise is still visible even after averaging over 200 realizations; the signal
would not be visible in the reconstructed image from a single noise realization.

The data domain ideal observer detectability is computed as a signal-to-noise ratio (SNR) for detection,
see Sec. 13.2.8 in Barrett and Myers\cite{barrett2004foundations},
which is straight-forward for the data model specified in Eqs. (\ref{datamean}) and (\ref{datavar}). For additive Gaussian noise,
using the small signal approximation, the ideal observer and ideal linear observer are equivalent.
The ideal linear observer performance is computed by first solving for the Hotelling template
\begin{linenomath}
\begin{equation*}
w_\text{data} = \frac{\bar{g}_\text{sig+} - \bar{g}_\text{sig--}}{\text{Var}(g_\text{sig--})},
\end{equation*}
\end{linenomath}
where 
\begin{linenomath}
\begin{align*}
\bar{g}_\text{sig+} &= P f_\text{sig+},\\
\bar{g}_\text{sig--} &= P f_\text{sig--};
\end{align*}
\end{linenomath}
and the small signal approximation is assuming
\begin{linenomath}
\begin{equation*}
\text{Var}(g_\text{sig+}) \approx
\text{Var}(g_\text{sig--}).
\end{equation*}
\end{linenomath}
The SNR for detection in the data domain is computed from the dot product of the Hotelling template and the signal projection data
\begin{linenomath}
\begin{equation*}
\text{SNR}^2_\text{data} = w_\text{data}^\top (\bar{g}_\text{sig+} - \bar{g}_\text{sig--}).
\end{equation*}
\end{linenomath}
The SNR metric can be converted to a receiver operating characteristic (ROC) area-under-the-curve (AUC), or equivalently
a percent-correct (PC) on a two-alternative-forced-choice (2-AFC) observer experiment (page 823 in Barrett and Myers\cite{barrett2004foundations})
\begin{linenomath}
\begin{equation*}
\text{PC}_\text{data} = \text{AUC}_\text{data} = \frac{1}{2} + \frac{1}{2} \text{erf} \left(\frac{\text{SNR}^2_\text{data}}{2}\right).
\end{equation*}
\end{linenomath}
For the equal-dose breast CT simulation at the specified noise level and the given signal properties, the signal detectability
in the data domain corresponds to
\begin{linenomath}
\begin{equation*}
\text{PC}_\text{data} = 86.57 \%,
\end{equation*}
\end{linenomath}
where the range of possible performance values are 50\%, corresponding to guessing on the 2-AFC experiment, to 100\%, a 2-AFC perfect
score. That the ideal observer performance is significantly less than 100\% in the data domain is intended by design.
This design requirement is why it is necessary to use the subtle signal shown in Fig. \ref{fig:detection}.

As pointed out in Sec. 13.2.6 of Barrett and Myers\cite{barrett2004foundations}, image reconstruction can only maintain or lose signal detectability with the ideal
observer, and as a result the ideal observer is not commonly used for assessing tomographic images after reconstruction.
Essentially, from the ideal observer perspective, image reconstruction should not be performed at all. Constrained by the fact
that human observers can interpret reconstructed images much more easily than sinograms, there is still potentially useful
knowledge to be gained from the ideal observer in assessing the efficiency of the image reconstruction algorithm; namely,
it can address the question of how well the separability between signal-present and signal-absent hypotheses is preserved
in passing through image reconstruction. In other words, does the image reconstruction algorithm wipe out the signal
in the detection task? This is a particularly relevant question for recent efforts in non-linear image reconstruction where
strong assumptions are being exploited to obtain tomographic images for sparse sampling conditions or low-dose scanning.
The image-domain ideal observer performance is also useful in that it provides a theoretical upper bound on human observer
performance, and no amount of post-processing will allow this bound to be exceeded.

For computing the image-domain detectability, we employ the 2-AFC PC figure-of-merit for the ideal observer in the image domain
because it is easy to interpret; the 2-AFC test intuitively connects image ensemble properties with single image noise realizations;
and we have a hard theoretical upper bound that it cannot exceed $\text{PC}_\text{data}=86.57$\%. This last property that,
\begin{linenomath}
\begin{equation*}
\text{PC}_\text{image} \le \text{PC}_\text{data},
\end{equation*}
\end{linenomath}
also naturally provides a measure for the loss of signal-detectability passing through image reconstruction.
To provide an accurate and precise estimate of $\text{PC}_\text{image}$, 4000 noisy data realizations of both signal-present
and signal-absent hypotheses are generated. All of the data realizations are reconstructed with the TV-LSQ algorithm.
Half of the resulting images under each hypothesis are used to train an ideal-observer classifier, and the remaining
half of the images is used to generate the $\text{PC}_\text{image}$ metric and its error bars.
(Because $\text{PC}_\text{image}$ is computed from noise realizations, it is necessary to work with a small signal 
due to its inherent uncertainty.
If the data domain PC is close to 100\%,
the resulting drop in going to the image domain PC may be too small to be significant.)
The large number of image realizations leads to a high precision, and the accuracy results from surveying a number
of classifiers including both ideal linear observer and ideal observer estimators. For the ideal linear observer, we have
investigated the channelized Hotelling observer \cite{gallas2003validating} with different channel formulations and a single-layer neural network (SLNN)
\cite{zhou2019approximating}. For the ideal observer, several implementations of a
convolutional neural network (CNN) \cite{zhou2019approximating} have been explored.
We have found that a hybrid-CHO yields $\text{PC}_\text{image}$ equal to the results, within error bars, from the NN classifiers
over the range of simulation parameters investigated.
We present the hybrid-CHO because of its relative simplicity, but the equivalence of the hybrid-CHO with the SLNN and CNN is significant
because the hybrid-CHO exploits approximate rotational symmetry in the detection task while the SLNN and CNN do not.
This approximate symmetry allows for a reduction in the number of channels needed for the hybrid-CHO, and the equivalence with the NN-based
observers indicates that the reduced set of channels in the hybrid-CHO is not compromising performance of the hybrid-CHO as an observer model.

\subsubsection{Hybrid-CHO}
The theory for estimation of the CHO $\text{PC}_\text{image}$ and its variance is covered in Gallas and Barrett\cite{gallas2003validating} and
Chen {\it et al.}\cite{chen2012classifier}.
The hybrid-CHO developed here exploits approximate rotational symmetry that
results from use of a small rotationally-symmetric signal and uniform angular sampling in the sinogram.
Because the detection task design is approximately rotationally symmetric, it 
lends itself well to the use of standard Laguerre-Gauss channels \cite{gallas2003validating}, which are circularly symmetric. 
The Laguerre-Gauss channels on their own, however, do not provide an optimal basis because of the small size of the signal
in combination with the fact that the image is discretized on a Cartesian grid. To account for both of these aspects of the
CT imaging set-up, we propose a hybrid channel set composed of Laguerre-Gauss channels combined with single-pixel channels
at the location of the signal. The observer model is referred to as a hybrid-CHO, reflecting this hybrid channel set.

The data for computing the hybrid-CHO performance consist of the central 128x128 region of pixels from each of the 512x512 image realizations;
thus there are a total of 4,000 signal-present and signal-absent 128x128 ROIs for training and testing the hybrid-CHO.
The continuous definition of the Laguerre-Gauss channels is
\begin{linenomath}
\begin{align}
\label{LG}
L_n &= \sum_{k=0}^n \binom{n}{k} \frac{(-1)^k}{k!} x^k \\
u_n(r|a) &= \frac{\sqrt{2}}{a} \exp \left( \frac{-\pi r^2}{a^2} \right) L_n \left(\frac{2 \pi r^2}{a^2} \right), \notag
\end{align}
\end{linenomath}
where the radius $r$ is defined $r^2=x^2+y^2$; $x,y$ indicate location on the 128x128 ROI; and the units of $x$ and $y$ are scaled
so that $(x,y)=(0,0)$ is the center of the ROI and $(x,y)=(1,1)$ is the upper right corner of the ROI. The parameters
of the Laguerre-Gauss channels are the order $n$ and Gaussian radial decay parameter $a$, which is specified in the same
scaled units as $r$.
The discrete representation of the Laguerre-Gauss channels is obtained by evaluating $u_n(r|a)$ at the center of each of the
pixels in the ROI.
The single-pixel channels are defined as
\begin{linenomath}
\begin{equation}
\label{SP}
u(s,t) =
\begin{cases}
1 & (i,j) = (s,t) \\
0 & (i,j) \neq (s,t)
\end{cases},
\end{equation}
\end{linenomath}
where $(i,j)$ are the integer coordinates of the pixels in 
the discrete channel function; $(s,t)$ is the location of the unit impulse;
the origin of the integer coordinates $(0,0)$ is at the lower left corner of the ROI.

The specific channel set employed for the breast CT simulation consists of fourteen channels.
The first ten are the discrete Laguerre-Gauss channels, $u_n(r|a)$, with $n\in [0,9]$ and $a=0.5$, and the remaining
for are the single-pixel channels, $\{ u(63,63),u(63,64),u(64,63),u(64,64) \}$.
Considering the channel functions as column vectors of length 128x128, where the pixel elements are in lexicographical order,
the 14 channels form a channelization matrix $U$ of size 16,384$\times$14 (16,384 = 128$\cdot$128).

The channelized linear classifier is computed by estimating the mean channelized signal and the channelized image
covariance. To compute these quantities, 
the channelized images are first obtained from the reconstructed training images by
\begin{linenomath}
\begin{align*}
[u_\text{sig+}]_i &= U^\top \left[ f^\text{(recon)}_\text{sig+}\right]_i,\\
[u_\text{sig--}]_i &= U^\top \left[ f^\text{(recon)}_\text{sig--}\right]_i,
\end{align*}
\end{linenomath}
where $i$ is the realization index, which runs from 1 to $N_\text{real}=4000$;
and $f^\text{(recon)}_\text{sig+}$ ($f^\text{(recon)}_\text{sig--}$) is a column vector with pixels values from
the central 128x128 ROI extracted from the
reconstruction from signal-present (signal-absent) data.
The first $i=1$ through $N_\text{train}=2000$ realizations are assigned to the training set,
and the rest of the realizations $i=N_\text{train}+1$ through $N_\text{train}+N_\text{test}$ are assigned
to the testing set.
The mean channelized signal is 
\begin{linenomath}
\begin{equation*}
s_u = (1/N_\text{train}) \sum_{i=1}^{N_\text{train}} ([u_\text{sig+}]_i -[u_\text{sig--}]_i).
\end{equation*}
\end{linenomath}
Using the small signal approximation, the training images under both hypotheses can be combined to provide
a covariance estimate
\begin{linenomath}
\begin{align*}
K_u = & (1/(2N_\text{train}-1)) \\
&  \left[\sum_{i=1}^{N_\text{train}}
([u_\text{sig+}]_i - \bar{u}_\text{sig+})^\top
([u_\text{sig+}]_i - \bar{u}_\text{sig+}) \right. \\
 + &  \left. \sum_{i=1}^{N_\text{train}}
([u_\text{sig--}]_i - \bar{u}_\text{sig--})^\top
([u_\text{sig--}]_i - \bar{u}_\text{sig--}) \right],
\end{align*}
\end{linenomath}
where the barred variables indicate mean over the ensemble of corresponding realizations.
The channelized Hotelling template is computed as
\begin{linenomath}
\begin{equation*}
w_u = K_u^{-1} s_u,
\end{equation*}
\end{linenomath}
and the ROI Hotelling template can be reconstituted by matrix-vector multiplication
\begin{linenomath}
\begin{equation*}
w_\text{image} = U w_u.
\end{equation*}
\end{linenomath}
Dotting a test image with the Hotelling template $w_\text{image}$ provides the test statistic,
which can be compared with a set threshold to make the classification into either signal-present
or signal-absent hypotheses.

The detectability metric in the image domain is estimated by running a 2-AFC experiment with
the hybrid-CHO for
every possible combination of signal-present and signal-absent test images
\begin{linenomath}
\begin{align*}
\label{pcimage}
\text{PC}_\text{image} &= (1/N_\text{test}^2) \sum_{i=1}^{N_\text{test}}\sum_{j=1}^{N_\text{test}}
c(a_i;b_j),\\
a_i &= w^\top_\text{image} \left[ f^\text{(recon)}_\text{sig+} \right]_{i+N_\text{train}}, \notag \\
b_j &= w^\top_\text{image} \left[ f^\text{(recon)}_\text{sig--} \right]_{j+N_\text{train}}, \notag
\end{align*}
\end{linenomath}
and the two-sample kernel function $c(a;b)$ is defined
\begin{linenomath}
\begin{equation*}
c(a;b) =
\begin{cases}
1 & a>b \\
0.5 &a=b \\
0 &a<b
\end{cases}.
\end{equation*}
\end{linenomath}
In the 2-AFC experiment, the Hotelling template is dotted with a pair of test images, where
one is drawn from the signal-present realizations and the other is drawn from the signal-absent realizations.
Whichever dot product yields the higher value, the hybrid-CHO classifies the corresponding image
as a signal-present image. The summation over the two-sample kernel function essentially counts all of the
times that the hybrid-CHO identified the signal-present image correctly.

Once $\text{PC}_\text{image}$ is computed it can be compared with $\text{PC}_\text{data}$ to provide a measure
of loss of signal detectability.  The quantity $\text{PC}_\text{data}$ is known analytically so the corresponding
value does not have error bars. The value $\text{PC}_\text{image}$, on the other hand, is estimated from realizations,
and thus it has variability due to the randomness of the testing set. There is also variability in the training
of the hybrid-CHO because it is computed from the random training images.  To account for both sources of variability
we employ the level 2 variability estimation from Chen {\it et al.}\cite{chen2012classifier}, and the 95\% confidence intervals are reported.

\subsection{Test phantom for visual correspondence}

\begin{figure}[!t]
\centerline{\includegraphics[width=0.5\columnwidth]{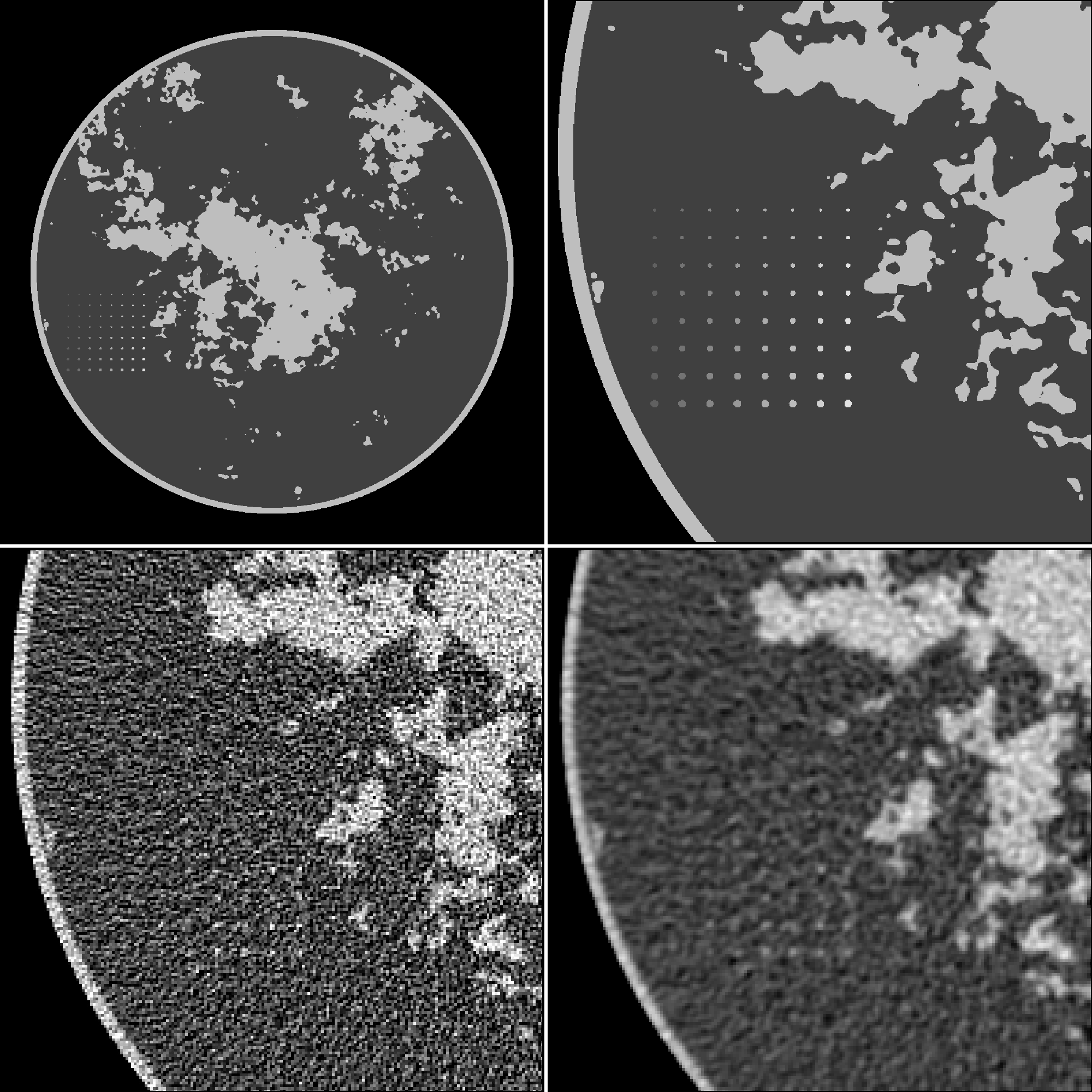}}
\caption{
Computerized breast phantom with a contrast-detail (CD) insert.
The displayed images are the image of the phantom (top, left), the ROI
focused on the CD insert (top, right), an unregularied FBP reconstruction
(bottom, left), and a regularized FBP image (bottom, right).
For reference, the RMSE values of the unregularized and regularized FBP images are 0.0198 and 0.01155 cm$^{-1}$,
respectively.
The gray scale window for all panels is [0.174,0.253]  cm$^{-1}$.}
\label{fig:phantom}
\end{figure}

In order to illustrate the correspondence between visual image quality and the image quality metrics,
the same simulation parameters and scan configurations are investigated using
a test phantom with a structured fibro-glandular tissue model \cite{Reiser10}, shown in Fig. \ref{fig:phantom}.
This breast phantom is composed of a 16 cm disk containing 
background fat tissue, attenuation 0.194 cm$^{-1}$, skin-line and randomly
generated fibro-glandular tissue at attenuation 0.233 cm$^{-1}$.
These components of the phantom are defined on a 2048$\times$2048 grid of dimensions
18$\times$18 cm$^{2}$.
The structured background allows for visualization of fine details. In order to have a more direct
comparison with a signal detection task, a contrast-detail (CD) insert is included in the phantom
consisting of an 8x8 grid of point-like signals.
The signals are defined as analytic disks so that
the line-integrals through the signals can be computed exactly and their
contribution to the projection data is not subject to pixelization of the test phantom image.
The disk contrast in the CD insert increases linearly from 0.01 cm$^{-1}$ to 0.05 cm$^{-1}$ going from left to right,
and the disk radius starts at 200 microns and increases linearly to 500 microns going from top to bottom.
For reference, the reconstruction grid's image pixel width is 350 microns.
To appreciate the noise level of the breast CT simulation, ROIs are shown of images reconstructed
by FBP using a ramp filter and FBP followed by Gaussian blurring.
For the FBP reconstructions, the $N_\text{views}=512$ scan configuration is used. Due to the high-level
of speckle noise in the unregularized FBP image, it is difficult to see even the most conspicuous
of signals in the CD insert. With regularization, the larger, higher contrast corner of the CD insert
becomes visible.

\section{Results}
\label{sec:results}

The hybrid-CHO signal detection figure-of-merit and image RMSE are computed alongside
TV-LSQ reconstructed images of the breast phantom, exploring the three parameters
of the CT-simulation: $N_\text{iter}$, $N_\text{views}$, and TV constraint parameter $\gamma$.
The TV constraint is reported as a fraction of the TV of the ground truth image.

\subsection{Signal detectability as a function of iteration number}

\begin{figure}[!t]
\centerline{\includegraphics[width=\columnwidth]{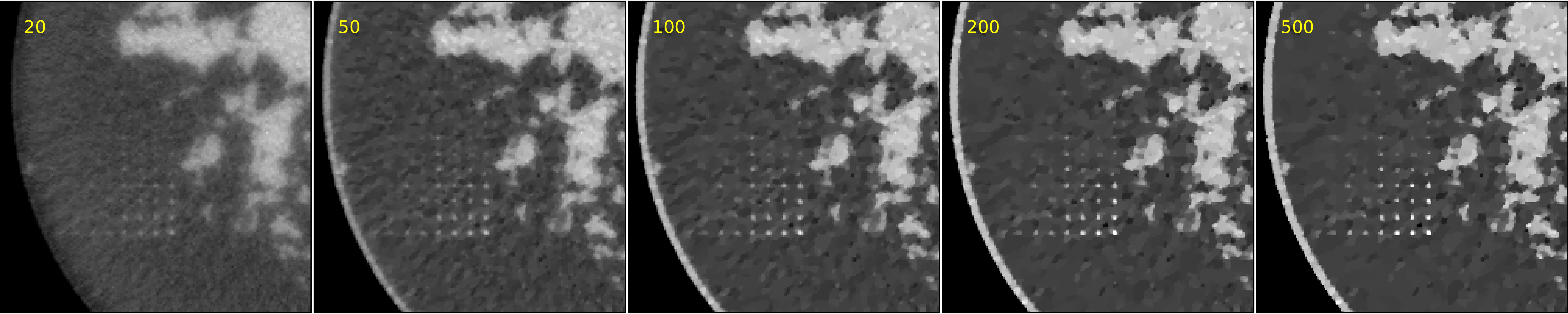}}
\centerline{~~~}
\centerline{\includegraphics[width=0.5\columnwidth]{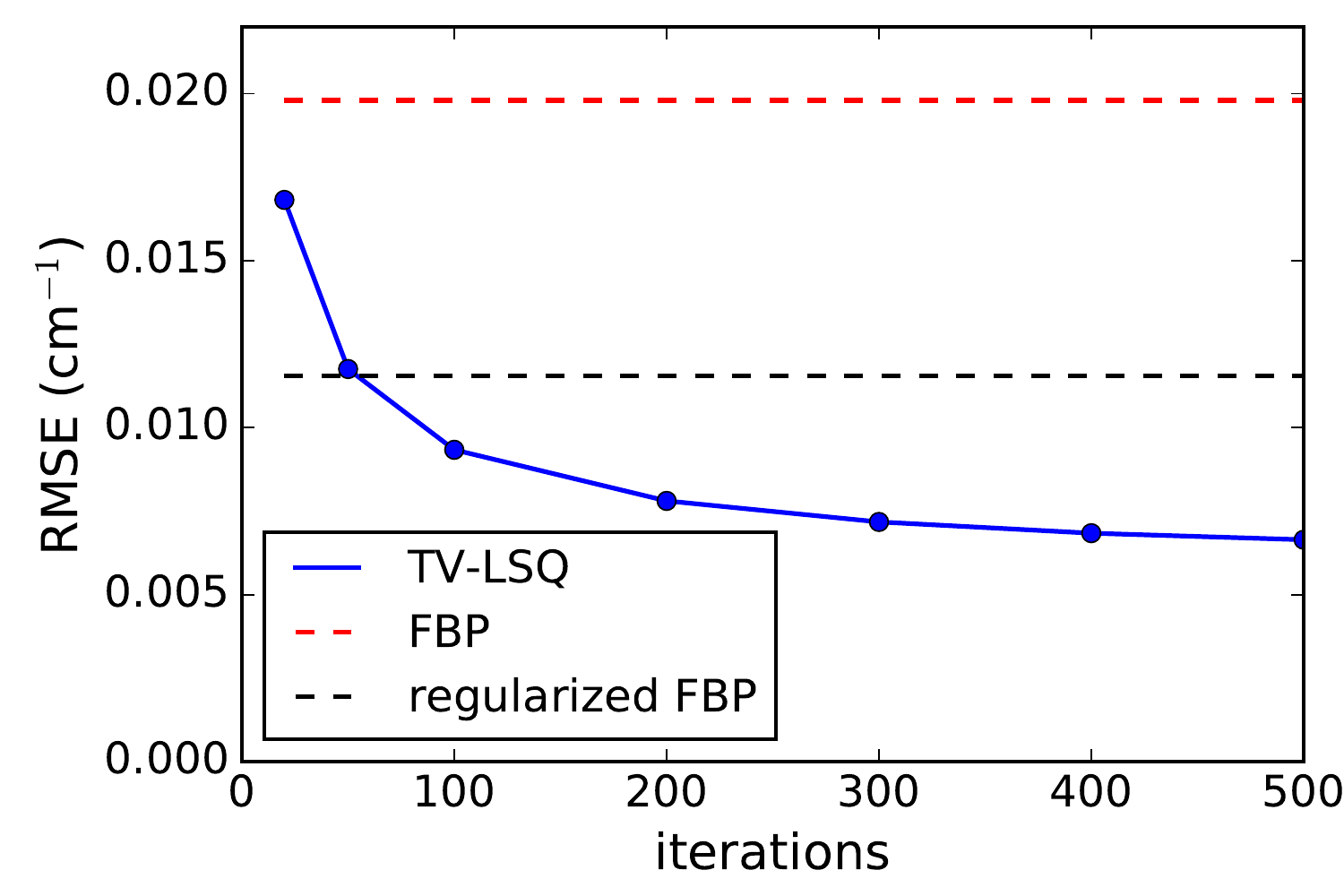}
\includegraphics[width=0.5\columnwidth]{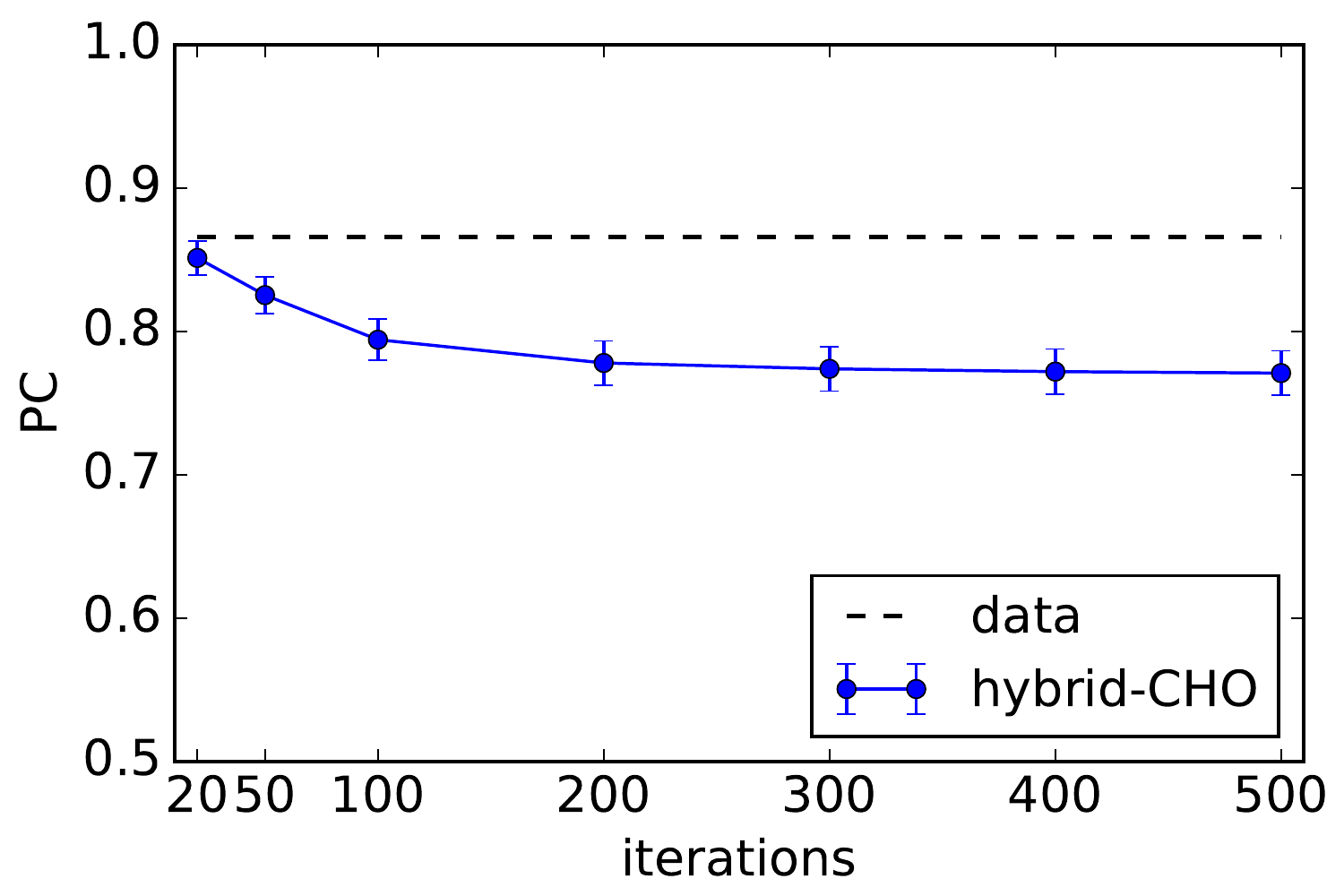}}
\caption{ (Top row)
Images reconstructed by TV-LSQ for $N_\text{views}=128$ and $\gamma=1.0$ with iteration number
increasing from left to right. The iteration number is indicated in each panel of the figure.
The gray scale window is [0.174,0.253]  cm$^{-1}$.
(Bottom, left) Plot of the corresponding image RMSE values.
For reference, the RMSE of the FBP and regularized FBP images from Fig. \ref{fig:phantom}
are 0.0198 and 0.01155 cm$^{-1}$, respectively. Both FBP values are indicated in the plot with dashed lines in red and black, respectively.
(Bottom, right) Plot of the corresponding signal detectability metric, percent correct for an ideal-observer 2-AFC experiment.
The dashed line indicates the theoretical
maximum PC performance inherent in the data domain; it does not depend on iteration number and is indicated
for reference.
}
\label{fig:iterresults}
\end{figure}

The first set of results focus on $N_\text{views}=128$ and $\gamma=1.0$, i.e. the TV constraint is equal
to the ground truth phantom TV.
A series of ROI images are shown in Fig. \ref{fig:iterresults} as a function of iteration number for the TV-LSQ
reconstruction of the breast phantom. From the perspective of accurate recovery of the phantom, the gray level
estimation appears to improve with increasing iteration number, as a general trend, which is to be expected
because the TV constraint is selected to be the TV of the test phantom. From the perspective of visualizing
the fine details in the image, the trend with iteration is more complex.
The structure detail in the fibro-glandular tissue and many of the signals are visible already
at 20 iterations, where it is clear from the overall gray value that the image is far from the solution
to the TV-LSQ problem.  As the iterations progress, the larger signals of the CD insert appear more conspicuous as the
speckle noise amplitude is reduced. On the other hand, some of the more subtle features in the image appear to become
distorted as the iterations progress, and the numerically converged image has a classic patchy look where it
is difficult to distinguish noise from real structures.

Corresponding to the image series in Fig. \ref{fig:iterresults}, quantitative image quality metrics
are also plotted, showing image RMSE and signal-detection
$\text{PC}_\text{image}$. As expected, the RMSE trend shows improvement with iteration number,
and the RMSE converges to a value well below that of the FBP reference images in Fig. \ref{fig:phantom}.
Again, $\gamma$ is set to the truth value and the test phantom has a high-degree of gradient sparsity;
thus the solution to the TV-LSQ optimization problem is expected to yield a mathematically accurate solution
and this is reflected in low RMSE values and the fact that the RMSE steadily improves as the TV-LSQ algorithm progresses
toward the solution. This trend coincides with the visual gray-level accuracy seen in the series of images.
It is interesting to note that the RMSE at $N_\text{iter}=500$ is substantially below the value of 0.01155 corresponding to the regularized
FBP image in Fig. \ref{fig:phantom}.

The iteration number trend for $\text{PC}_\text{image}$, however, runs opposite to the
image RMSE. There is a clear decline in the signal detectability at early iterations, and as
convergence is achieved this metric plateaus to a value well below the data domain signal detectability.
The trend in image detectability coincides with the visual appearance of the
the increasing patchiness of the images shown
in Fig. \ref{fig:iterresults}. 

The main point of the $\text{PC}_\text{image}$ metric is that it should reflect the disappearance of small subtle
details in the image, and in this example we see correspondence between this metric 
and the overall patchiness of the images.
Thus the quantitative $\text{PC}_\text{image}$ metric appears to capture
the desired image properties, providing a quantitative measure of over-regularization.
How to use this information to determine algorithm parameters depends on the goal of the CT system design.
Clearly, the results of the iteration number study indicate that $\text{PC}_\text{image}$ cannot be used alone
to determine the optimal iteration number, because it has the largest value with one iteration.
As an aside, we note that a similar behavior was observed for the maximum likelihood expectation maximization (MLEM) algorithm
using a ROI-observer \cite{abbey1996observer}, and we take up a comparison of these experiments in Sec. \ref{sec:discussion}.

Using $\text{PC}_\text{image}$ in concert with image RMSE, which has the opposite trend, provides complimentary information.
As an example of how it can be used, the desired
image could be specified by minimizing RMSE with the constraint that the loss in signal detectability
is bounded by a parameter $\epsilon$, i.e. $\text{PC}_\text{image}/\text{PC}_\text{data} \ge \epsilon$.
Subjectively, the first frame at 20 iterations, in the series of images shown in Fig. \ref{fig:iterresults},
has the best visibility for the signals in the CD insert and image texture realism.
The next image at 50 iterations already has a patchy appearance. Visualization of the intermediate frames (not shown)
suggests that a value of $\epsilon = 97.0$\% for this particular example provides a useful bound.
However, the details of how $\epsilon$ is chosen and how the detection task is designed depends on the
desired imaging goal. Here, we only aim to establish correspondence between $\text{PC}_\text{image}$
and the subjective image quality of patchiness or over-regularization with non-linear image reconstruction.

\subsection{Signal detectability as a function of $N_\text{views}$ and $\gamma$}

\begin{figure}[!t]
\centerline{\includegraphics[width=\columnwidth]{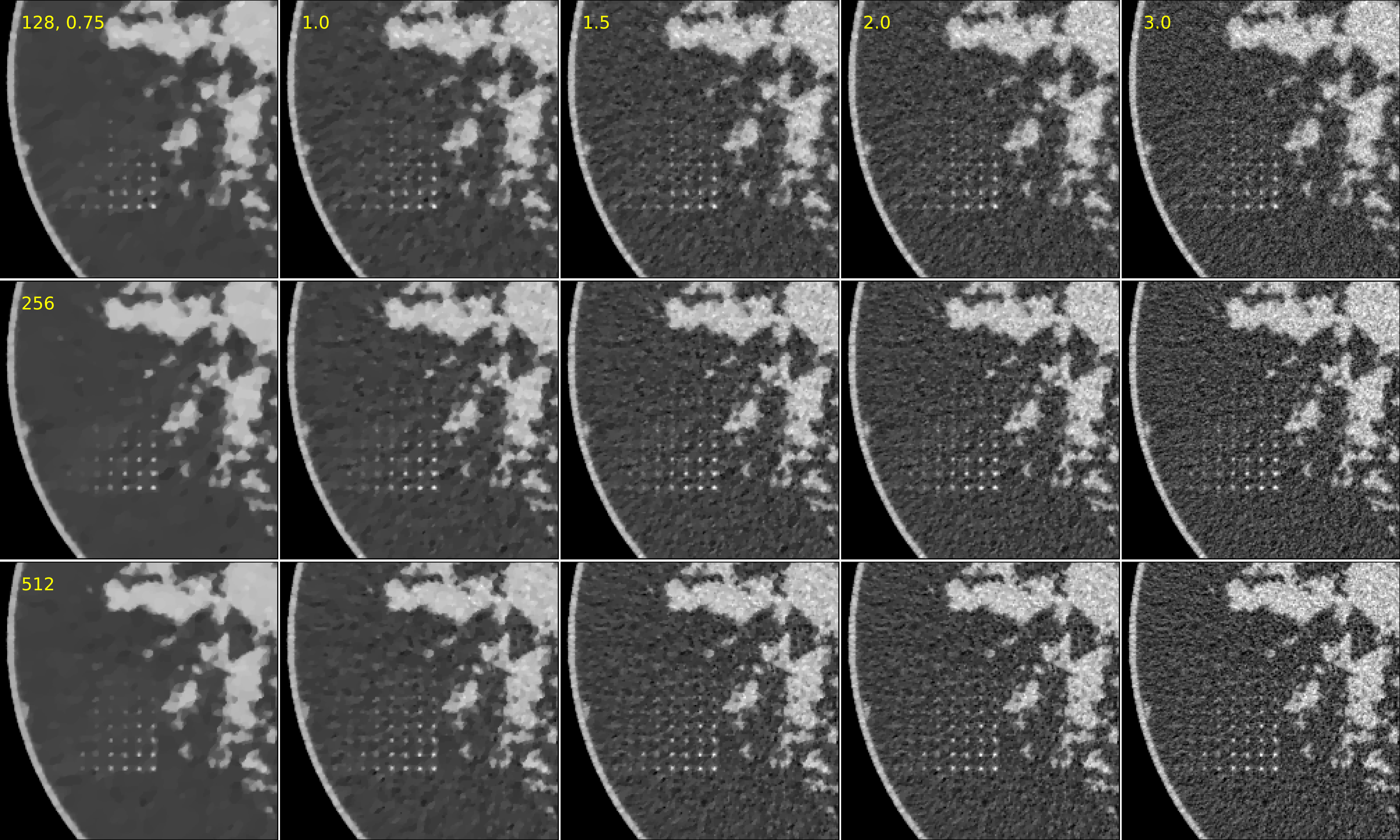}}
\centerline{~~~}
\centerline{\includegraphics[width=0.5\columnwidth]{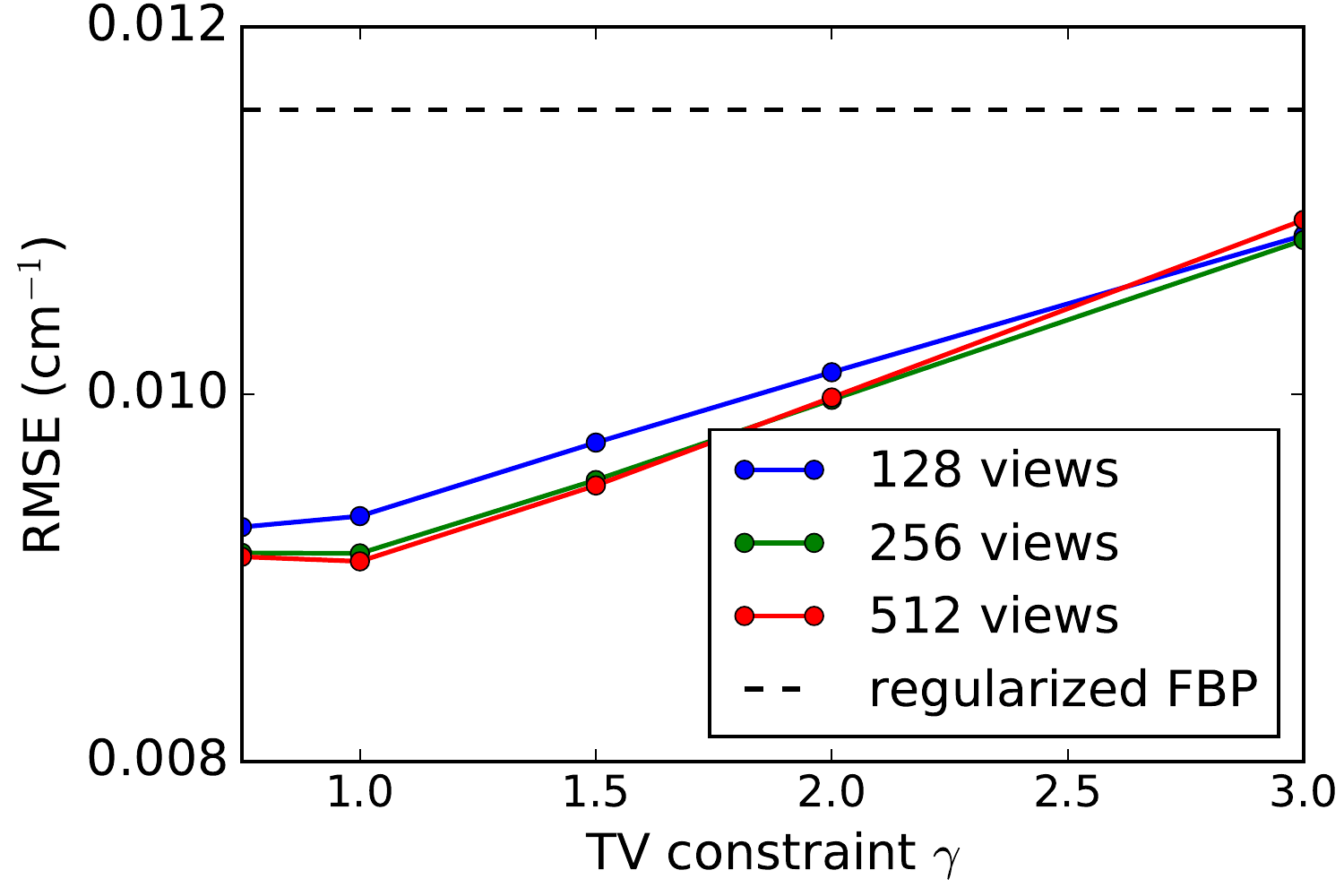}
\includegraphics[width=0.5\columnwidth]{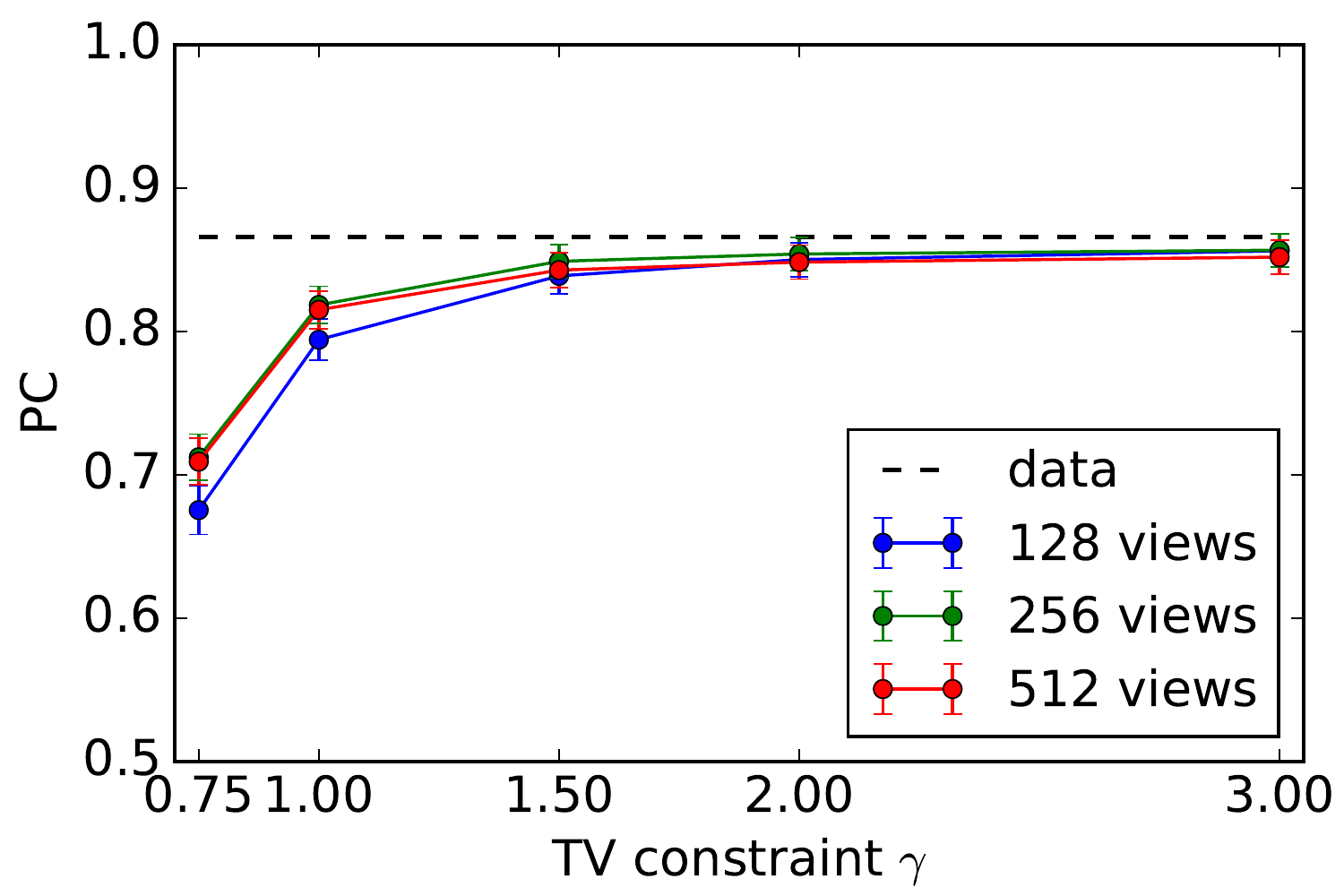}}
\caption{(Top row)
Images reconstructed by TV-LSQ for $N_\text{iter}=100$, varying $N_\text{views}$ from top to
bottom and varying $\gamma$ from left to right.
These parameters are indicated in the figure panels.
The gray scale window is [0.174,0.253]  cm$^{-1}$.
(Bottom, left) Plot of the corresponding image RMSE values.
For reference, the RMSE of the FBP and regularized FBP images from Fig. \ref{fig:phantom}
are 0.0198 and 0.01155 cm$^{-1}$, respectively, and the latter value is indicated in the plot with a dashed line.
(Bottom, right) Plot of the corresponding signal detectability metric, percent correct for an ideal-observer 2-AFC experiment.
The dashed line indicates the theoretical
maximum PC performance inherent in the data domain.
}
\label{fig:gridresults}
\end{figure}

For the next set of results, we fix $N_\text{iter}=100$ and vary the other two parameters of the breast CT simulation.
In Fig. \ref{fig:gridresults}, a grid of images is shown with each row and column corresponding to fixed $N_\text{views}$ and
$\gamma$, respectively.
As a general trend the lower $\gamma$ values reduce the speckle noise and streaks in the image; however, it is also
clear that the heavy regularization imposed by $\gamma=0.75$ effectively renders the borderline signals in the CD insert invisible.
In terms of conspicuity of the signals in the CD insert, the images for $\gamma=1.5$ and above appear to have similar numbers of signals visible.
The trend in $N_\text{views}$ is more difficult to discern because the conditions of the scan are set up to be equal dose.
For the larger $\gamma$-values $N_\text{views}=128$ images appear to have streak artifacts in addition to the speckle noise.
In general, there is a different noise texture for the various equal-dose scan configurations.

The corresponding image RMSE and $\text{PC}_\text{image}$ IQ metrics are also  plotted in Fig. \ref{fig:gridresults}.
The image RMSE favors $\gamma=1.0$, the ground truth TV value, although the RMSE for $\gamma=0.75$ is only slightly larger.
Also, the RMSE values decrease weakly with increasing $N_\text{views}$. The $\text{PC}_\text{image}$ values favor an opposite
trend, where the signal detectability increases with $\gamma$. Interestingly, for the different $N_\text{views}$ configurations,
the intermediate value $N_\text{views}=256$ is slightly favored, although the values for 256 and 512
have overlapping error bars.

Again, we point out that the metrics are complimentary.
Going by $\text{PC}_\text{image}$ alone the TV constraint would be abandoned.
Going by image RMSE alone, however, can also lead to an equally pathological situation where the image is over-regularized.
Using $\text{PC}_\text{image}$ in concert with RMSE
yields a more useful picture. We observe that, while it is true that $\text{PC}_\text{image}$ is monotonically increasing with $\gamma$,
there is clearly diminishing returns for $\gamma \ge 1.5$, where this metric appears to plateau. The RMSE, on the other hand,
favors lower $\gamma$ on the $\text{PC}_\text{image}$-plateau. Thus a prescription that combines the two metrics could
reasonably select an intermediate $\gamma$ value such as $\gamma=1.5$, where again 
$\epsilon = \text{PC}_\text{image}/\text{PC}_\text{data} \ge 97$\%.
At this setting, we observe that the TV-LSQ reconstructed images in Fig. \ref{fig:gridresults} do not have the patchy appearance of
over-regularization with TV. Also, compared with the FBP images, the image RMSE is lower and more CD insert signals are visible for TV-LSQ at $\gamma=1.5$.

\subsection{Estimation of subject TV and its impact on IQ metric trends}

The dependence of the simulation results on $\gamma$ are all referred to the ground truth TV value, which is
object dependent. Thus applying the simulation-based IQ metrics to an actual scanning situation,
where the ground truth is unknown, raises two important questions: (1) how to determine the subject TV, and
(2) does the subject TV reference value yield universal IQ metric dependence on $\gamma$. Two simulations
are performed to address both of these questions.

\begin{figure}[!t]
\centerline{\includegraphics[width=0.5\columnwidth]{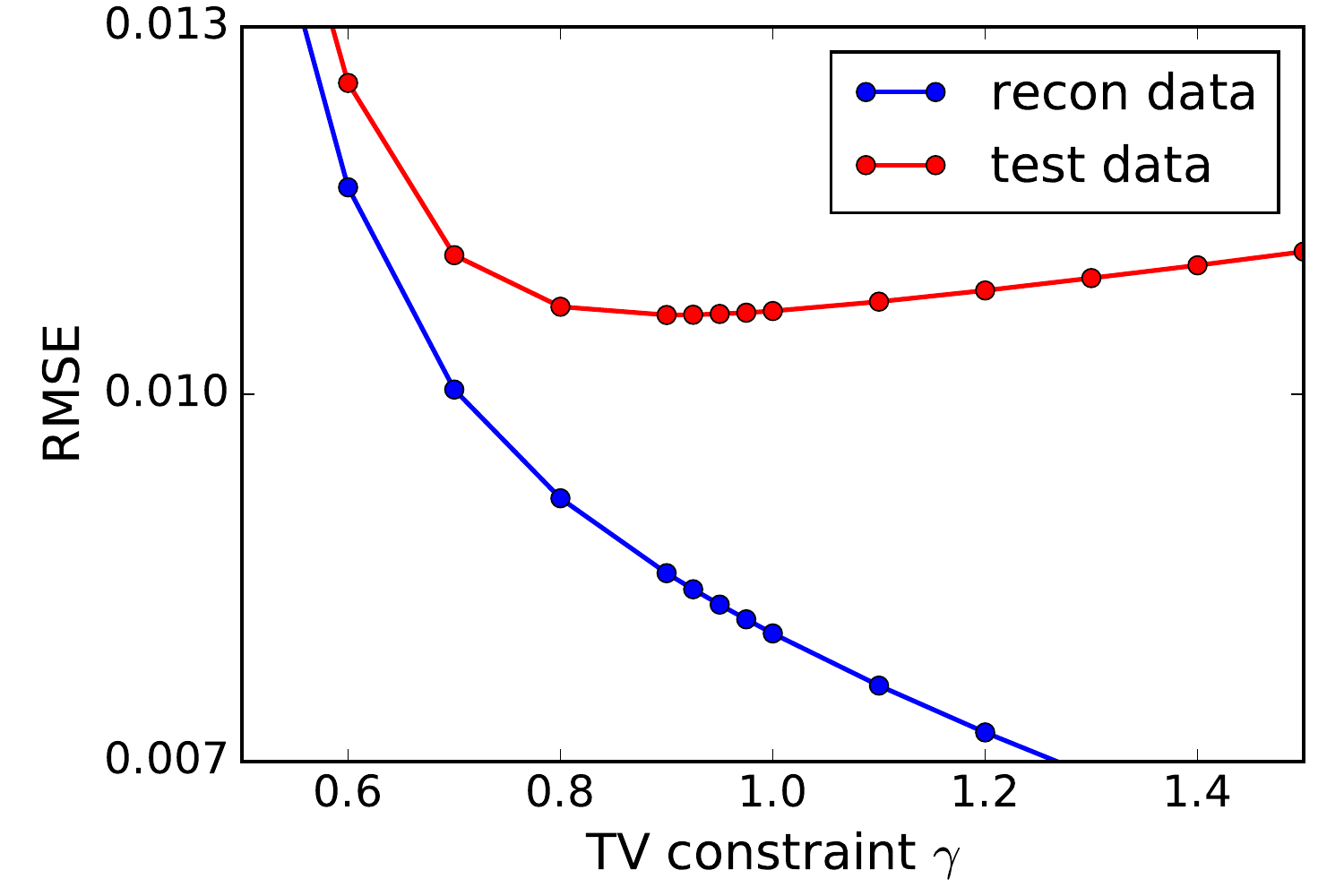}}
\caption{Plot of RMSE for the data used in image reconstruction (blue) and the RMSE on the left-out testing
data used for validation (red) set as a function
of the TV constraint parameter. The validation RMSE has a minimum at $\gamma=0.9$ in units scaled to
the ground truth image TV.
}
\label{fig:validation}
\end{figure}

To estimate
the subject TV, $\gamma_0$, we have successfully applied a validation technique \cite{schmidt2017spectral} where
image reconstruction is performed with a fraction of the available data and the remaining test data
are compared with the projection of the estimated image. The constraint value is estimated to be the value
that yields the smallest discrepancy between the test data and the corresponding estimated data.
We perform this computation in the context of the
present breast CT simulation for $N_\text{views}=128$ and $N_\text{iter}=500$. Image reconstruction is performed with 90\% of the
available line-integration data, chosen from the sinogram at random. This leaves 10\% of the data for independent testing.
The resulting reconstructed image is projected and the RMSEs
for the reconstruction and testing data are plotted in Fig. \ref{fig:validation} as a function of $\gamma$.
From Fig. \ref{fig:validation}, we observe that there is a monotonically decreasing trend in the reconstruction data RMSE
as a function of $\gamma$, but the data RMSE of the testing set shows a minimum at $\gamma=0.9$, which is close to the
true value of $\gamma=1.0$.
This result demonstrates that this validation technique can provide
an estimate of the subject TV to within 10 percent.

\begin{figure}[!t]
\centerline{\includegraphics[width=0.5\columnwidth]{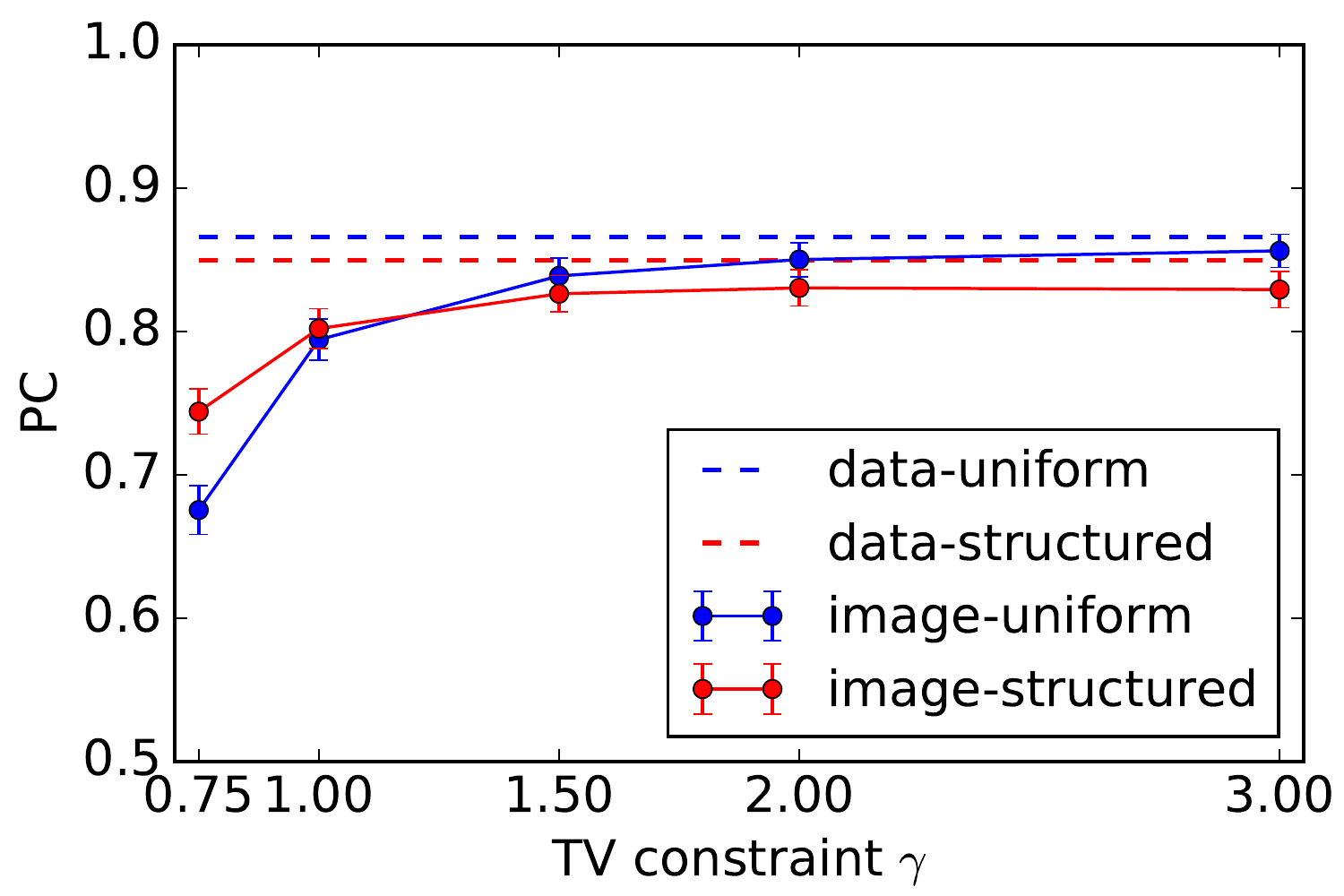}}
\caption{Detectability metrics using different background images. The label ``uniform''
and refers to the use of the
background image shown in Fig. \ref{fig:detection}, and ``structured'' refers to using the breast
phantom in Fig. \ref{fig:phantom} as the background image. Note that the data domain percent correct
is lower for the structured background because it is more attenuating.
}
\label{fig:ucho}
\end{figure}

To address the universality question, the uniform background used in the
process of estimating $\text{PC}_\text{image}$ is changed to the non-uniform, but known, background of the
breast phantom. This modification alters $\gamma_0$ dramatically; thus it is of interest to compare
the resulting $\text{PC}_\text{image}$ curve as a function of $\gamma$. In Fig. \ref{fig:ucho} this metric
is plotted for $N_\text{views}=128$ and $N_\text{iter}=100$.
From the graph it is clear that there is some numerical discrepancy between the numerical values
of $\text{PC}_\text{image}$ for the same value of the scaled parameter $\gamma$; however, the trend of this
metric as a function of $\gamma$ matches fairly well. That there is discrepancy in the
absolute numerical values is perhaps not too surprising considering the large difference in
background structure. The similarity in trends is further evidence of the potential utility of the
proposed IQ metric.

\section{Discussion}
\label{sec:discussion}

The proposed signal detectability index for non-linear image reconstruction bears some similarity
with the signal detectability studies on MLEM iteration number studies presented by Abbey {\it et al.}\cite{abbey1996observer}. 
In particular, the ROI observer from that investigation showed a steadily decreasing trend with iteration number.
The two detectability indices, however, are different and need to be interpreted differently.
The detection task considered in Abbey {\it et al.}\cite{abbey1996observer} was meant to have direct relevance to a clinical detection
task, and furthermore the authors were seeking correspondence between model and human observers on signal detection.
For the detectability metric, presented here, the signal size and amplitude are chosen so that the signal
is on the edge of detectability by the ideal observer in the data space. This signal is much too small to be detected
by a human observer; thus the detection task design itself is not directly relevant to a clinical detection task.
The design and purpose of this detection task is meant to be a surrogate for the subjective image
property of patchiness specific to over-regularization in TV-LSQ reconstructed images. 

The reduction of $\text{PC}_\text{image}$ relative to $\text{PC}_\text{data}$ represents an irretrievable loss
of information in distinguishing signal-present and signal-absent hypothesis. No post-processing operations
can improve on $\text{PC}_\text{image}$. This metric, however, only captures loss of detectability
due to non-invertibility of the image reconstruction algorithm. It does not necessarily reflect distortion of the
signal. For example, regularizing FBP images with moderate blurring, such as what is seen in Fig. \ref{fig:phantom},
is invertible and does not cause a reduction in PC even though the signal itself is broadened by the blurring operation.
Reconstructing FBP images onto an image grid of large pixels, on the other hand, is a non-invertible and does cause
loss in detectability \cite{sanchez2013comparison}.

The SKE/BKE detection task paradigm with a small rotationally-symmetric signal and uniform projection-angle sampling
allows for the hybrid-CHO to accurately represent the ideal linear observer with a relatively small set of channels,
because the detection task is well-suited to the rotationally-symmetric LG channels. Considering non-rotationally symmetric
signals or scanning angular ranges less than 2$\pi$ breaks this symmetry. In such cases, a different channel
representation and possibly a larger channel set would need to be developed in order for the hybrid-CHO to represent
the ideal linear observer. The SKE/BKE detection task also considers the signal at one location in the image.
For the presented non-linear image reconstruction algorithm, this limitation does not impact the utility of the
metric because the TV regularization is applied isotropically over the image and results are not expected
to change appreciably for different signal locations. Regularization techniques that involve
spatially varying weighting need to consider either multiple SKE/BKE detection tasks with signals at different locations
or a signal-known-statistically (SKS) detection task where the signal location is drawn from a spatially uniform
probability distribution.

\section{Conclusion}
\label{sec:conclusion}

We have developed and presented an image quality metric that is sensitive to the removal of subtle details
in the image and that can be applied to the non-linear TV-LSQ image reconstruction algorithm. The metric
is based on the detection of a small signal at the border of detectability by the ideal observer, and this
metric is hypothesized to quantify 
the subjective visual removal of subtle image details.
The design of the proposed detection task, use of the ideal observer, and connection with the 2-AFC
experiment makes the metric easy to interpret.
The detectability index, which is an estimate of a property
of an ensemble of reconstructed images, is connected to single image realizations through the interpretation as
a PC on a 2-AFC experiment.
Loss of detectability through the image reconstruction process, i.e. $\text{PC}_\text{image} < \text{PC}_\text{data}$,
unambiguously represents a quantitative decrease in the ability to distinguish signal-present and signal-absent images.
The bounds on this metric are clear: $0.5 \le \text{PC}_\text{image} \le \text{PC}_\text{data}$, where the lower
limit of 0.5 represents guessing on the 2-AFC experiment and the upper limit is the analytically known $\text{PC}_\text{data}$.

Correspondence between $\text{PC}_\text{image}$ and visual assessment of the reconstructed images of the
breast CT simulation shows that this metric may serve to quantify TV-LSQ over-regularization.
A decrease in this metric is seen to coincide with loss of borderline signals in the CD insert and with
patchiness in the appearance of the images.
This metric is seen to be complimentary to widely used image fidelity metrics such as image RMSE, and it may
help to provide an objective means to establish useful tomographic system parameter settings.
The presented methodology may also prove useful for quantifying over-regularization with other non-linear image
reconstruction techniques.

\appendix

\section{Appendix: The CPPD algorithm for TV-LSQ}
\label{app:cppd}

The CPPD algorithm \cite{chambolle2011first,Pock2011} can be used to efficiently 
solve non-smooth convex optimization problems for CT image reconstruction \cite{SidkyCP:12}.
We provide the pseudocode for CPPD-TV-LSQ in Algorithm \ref{alg:cppd}.

\begin{algorithm}
\hrulefill
\begin{algorithmic}[1]
\State $k \gets 0$, $f^{(0)} \gets 0$, $\lambda_s^{(0)} \gets 0$,  $\lambda_g^{(0)} \gets 0$ 
\While{$k < N_\text{iter}$}
\State $f^{(k+1)} \gets f^{(k)} - \tau \left( \nu_s X^\top \lambda_s^{(k)} + \nu_g D^\top \lambda_g^{(k)} \right)$
\State $\bar{f} \gets 2f^{(k+1)} -f^{(k)}$
\State $\lambda_s^{(k+1)} \gets \left( \lambda_s^{(k)} + \sigma(\nu_s X\bar{f} - \nu_s g)\right)/ (1+\sigma)$
\State $\lambda_g^+ \gets \lambda_g^{(k)} + \sigma \nu_g D \bar{f}$
\State $p \gets |\lambda_g^+|_\text{mag}$
\If {$\|p\|_1 > \nu_g \gamma \sigma $}
\State $\beta^{(k+1)} \gets \solve(\beta,\|\sh(p,\beta) \|_1 - \nu_g \gamma\sigma  =0) $
\State $\lambda_g^{(k+1)} \gets  \beta^{(k+1)}  \lambda_g^+ / \max \left( \beta^{(k+1)}, p \right)$
\Else
\State $\beta^{(k+1)} \gets 0$
\State $\lambda_g^{(k+1)} \gets  0$
\EndIf
\State $k \gets k+1$
\EndWhile
\end{algorithmic}
\hrulefill
\caption{
Pseudocode for the CPPD-TV-LSQ algorithm. The only free algorithm parameters are $\rho$ and
$N_\text{iter}$. Other algorithm inputs are the sinogram data $g$ and the TV constraint parameter $\gamma$.
The algorithm parameters and operations are explained in the text.}
\label{alg:cppd}
\end{algorithm}

In the CPPD-TV-LSQ algorithm, the parameters $\nu_s$ and $\nu_g$ normalize the
linear transforms $X$ and $D$:
\begin{equation*}
\nu_s = 1/\|X\|_2, \; \;
\nu_g = 1/\|D\|_2 ,
\end{equation*}
where the $\ell_2$-norm of a matrix is its largest singular value.
This scaling is performed so that algorithm efficiency is optimized and so that results
are independent of the physical units used in implementing $X$ and $D$. Note
that the sinogram $g$ in line 5 and the TV constraint parameter $\gamma$ in lines 8 and 9
must also be multiplied
by $\nu_s$ and $\nu_g$, respectively.
The step-size parameters $\sigma$ and $\tau$ must satisfy the inequality
\begin{equation}
\label{sineq}
\sigma \tau \le 1/L^2,
\end{equation}
where $L$ is the matrix norm of $A$, which is 
constructed by stacking $\nu_s X$ on $\nu_g D$
\begin{equation*}
A= \binom{\nu_s X}{\nu_g D}, \; \;
L= \| A \|_2.
\end{equation*}
Due the normalization of $X$ and $D$ and the fact that $X$ and $D$ approximately commute,
$L$ should be close to 1.
The step-size inequality, Eq. (\ref{sineq}), is satisfied with equality by setting
\begin{equation*}
\sigma = \rho/L,\;\;
\tau= 1/(\rho L),
\end{equation*}
where the step-size ratio $\rho$ is a free parameter that must be tuned
because it can strongly impact CPPD convergence behavior.
For all the simulations presented in this work the step-size ratio is set to $\rho=1$.

The ``solve'' function at line 9, returns the value of $\beta$ that
solves the equation written in its second argument, and ``shrink'' is defined
component-wise as
\begin{equation*}
[\text{shrink}(p,\beta)]_i =
\begin{cases}
p_i +\beta & p_i < -\beta \\
0          & |p_i| \le \beta\\
p_i-\beta  & p_i > \beta
\end{cases} ,
\end{equation*}
where $i$ is an index for the components of $p$.
Solution of the equation at line 9 can be implemented by bisection, because
$\|\text{shrink}(p,\beta)\|_1$ decreases monotonically as $\beta$ increases and
the root of the equation is bracketed in the
interval $[0,\max(p)]$, where $\max$ acts component-wise on $p$.

\section*{Acknowledgment}
This work is supported in part by NIH
Grant Nos. R01-EB026282, R01-EB023968, R01-EB020604, R01-EB028652
and The University of Chicago Women's Board. The computational resources
for this work is funded in part by the NIH S10-OD025081, S10-RR021039, and P30-CA14599 awards.
The contents of this article are solely the responsibility of
the authors and do not necessarily represent the official
views of the National Institutes of Health.


\begin{thebibliography}{10}

\bibitem{elbakri2002statistical}
I.~A. Elbakri and J.~A. Fessler,
\newblock Statistical image reconstruction for polyenergetic {X}-ray computed
  tomography,
\newblock IEEE Trans. Med. Imaging {\bf 21}, 89--99 (2002).

\bibitem{mccollough2009strategies}
C.~H. McCollough, A.~N. Primak, N.~Braun, J.~Kofler, L.~Yu, and J.~Christner,
\newblock Strategies for reducing radiation dose in {CT},
\newblock Radiol. Clinics {\bf 47}, 27--40 (2009).

\bibitem{sidky2008image}
E.~Y. Sidky and X.~Pan,
\newblock Image reconstruction in circular cone-beam computed tomography by
  constrained, total-variation minimization,
\newblock Phys. Med. Biol. {\bf 53}, 4777--4807 (2008).

\bibitem{chen2008prior}
G.-H. Chen, J.~Tang, and S.~Leng,
\newblock Prior image constrained compressed sensing {(PICCS)}: a method to
  accurately reconstruct dynamic {CT} images from highly undersampled
  projection data sets,
\newblock Med. Phys. {\bf 35}, 660--663 (2008).

\bibitem{ritschl2011improved}
L.~Ritschl, F.~Bergner, C.~Fleischmann, and M.~Kachelrie{\ss},
\newblock Improved total variation-based {CT} image reconstruction applied to
  clinical data,
\newblock Phys. Med. Biol. {\bf 56}, 1545 (2011).

\bibitem{batenburg2011dart}
K.~J. Batenburg and J.~Sijbers,
\newblock {DART}: a practical reconstruction algorithm for discrete tomography,
\newblock IEEE Trans. Image Proc. {\bf 20}, 2542--2553 (2011).

\bibitem{gupta2018cnn}
H.~Gupta, K.~H. Jin, H.~Q. Nguyen, M.~T. McCann, and M.~Unser,
\newblock {CNN}-based projected gradient descent for consistent {CT} image
  reconstruction,
\newblock IEEE Trans. Med. Imaging {\bf 37}, 1440--1453 (2018).

\bibitem{adler2018learned}
J.~Adler and O.~{\"O}ktem,
\newblock Learned primal-dual reconstruction,
\newblock IEEE Trans Med. Imaging {\bf 37}, 1322--1332 (2018).

\bibitem{barrett2004foundations}
H.~H. Barrett and K.~J. Myers,
\newblock {\em Foundations of image science},
\newblock Wiley, Hoboken, New Jersey, 2004.

\bibitem{abbey1996observer}
C.~K. Abbey, H.~H. Barrett, and D.~W. Wilson,
\newblock Observer signal-to-noise ratios for the {ML-EM} algorithm,
\newblock in {\em Medical Imaging: Image Perception}, edited by H.~L. Kundel,
  volume 2712, pages 47--58, Proc. SPIE, 1996.

\bibitem{abbey2001human}
C.~K. Abbey and H.~H. Barrett,
\newblock Human-and model-observer performance in ramp-spectrum noise: effects
  of regularization and object variability,
\newblock JOSA A {\bf 18}, 473--488 (2001).

\bibitem{wunderlich2008image}
A.~Wunderlich and F.~Noo,
\newblock Image covariance and lesion detectability in direct fan-beam x-ray
  computed tomography,
\newblock Phys. Med. Biol. {\bf 53}, 2471--2493 (2008).

\bibitem{das2010penalized}
M.~Das, H.~C. Gifford, J.~M. O'Connor, and S.~J. Glick,
\newblock Penalized maximum likelihood reconstruction for improved
  microcalcification detection in breast tomosynthesis,
\newblock IEEE Transactions on Medical Imaging {\bf 30}, 904--914 (2010).

\bibitem{sanchez2013comparison}
A.~A. Sanchez, E.~Y. Sidky, I.~Reiser, and X.~Pan,
\newblock Comparison of human and Hotelling observer performance for a fan-beam
  {CT} signal detection task,
\newblock Med. Phys. {\bf 40}, 031104 (2013).

\bibitem{gang2014task}
G.~J. Gang, J.~W. Stayman, W.~Zbijewski, and J.~H. Siewerdsen,
\newblock Task-based detectability in {CT} image reconstruction by filtered
  backprojection and penalized likelihood estimation,
\newblock Med. Phys. {\bf 41}, 081902 (2014).

\bibitem{sanchez2014task}
A.~A. Sanchez, E.~Y. Sidky, and X.~Pan,
\newblock Task-based optimization of dedicated breast {CT} via {H}otelling
  observer metrics,
\newblock Med. Phys. {\bf 41}, 101917 (2014).

\bibitem{xu2015task}
J.~Xu, M.~K. Fuld, G.~S.~K. Fung, and B.~M.~W. Tsui,
\newblock Task-based image quality evaluation of iterative reconstruction
  methods for low dose CT using computer simulations,
\newblock Phys. Med. Biol. {\bf 60}, 2881--2901 (2015).

\bibitem{rose2017investigating}
S.~D. Rose, A.~A. Sanchez, E.~Y. Sidky, and X.~Pan,
\newblock Investigating simulation-based metrics for characterizing linear
  iterative reconstruction in digital breast tomosynthesis,
\newblock Med. Phys. {\bf 44}, e279--e296 (2017).

\bibitem{liu2013total}
Y.~Liu, Z.~Liang, J.~Ma, H.~Lu, K.~Wang, H.~Zhang, and W.~Moore,
\newblock Total variation-stokes strategy for sparse-view X-ray CT image
  reconstruction,
\newblock IEEE Trans. Med. Imaging {\bf 33}, 749--763 (2013).

\bibitem{niu2014sparse}
S.~Niu, Y.~Gao, Z.~Bian, J.~Huang, W.~Chen, G.~Yu, Z.~Liang, and J.~Ma,
\newblock Sparse-view x-ray CT reconstruction via total generalized variation
  regularization,
\newblock Phys. Med. Biol. {\bf 59}, 2997--3017 (2014).

\bibitem{boone2005technique}
J.~M. Boone, A.~L.~C. Kwan, J.~A. Seibert, N.~Shah, K.~K. Lindfors, and T.~R.
  Nelson,
\newblock Technique factors and their relationship to radiation dose in pendant
  geometry breast {CT},
\newblock Med. Phys. {\bf 32}, 3767--3776 (2005).

\bibitem{jorgensen2015little}
J.~S. J{\o}rgensen and E.~Y. Sidky,
\newblock How little data is enough? {P}hase-diagram analysis of
  sparsity-regularized {X}-ray computed tomography,
\newblock Philos. Trans. Royal Soc. {A} {\bf 373}, 20140387 (2015).

\bibitem{chambolle2011first}
A.~Chambolle and T.~Pock,
\newblock A first-order primal-dual algorithm for convex problems with
  applications to imaging,
\newblock J. Math. Imag. Vis. {\bf 40}, 120--145 (2011).

\bibitem{Pock2011}
T.~Pock and A.~Chambolle,
\newblock Diagonal preconditioning for first order primal-dual algorithms in
  convex optimization,
\newblock in {\em International Conference on Computer Vision ({ICCV} 2011)},
  pages 1762--1769, 2011.

\bibitem{SidkyCP:12}
E.~Y. Sidky, J.~H. J{\o}rgensen, and X.~Pan,
\newblock Convex optimization problem prototyping for image reconstruction in
  computed tomography with the {C}hambolle-{P}ock algorithm,
\newblock Phys. Med. Biol. {\bf 57}, 3065--3091 (2012).

\bibitem{gallas2003validating}
B.~D. Gallas and H.~H. Barrett,
\newblock Validating the use of channels to estimate the ideal linear observer,
\newblock JOSA A {\bf 20}, 1725--1738 (2003).

\bibitem{zhou2019approximating}
W.~Zhou, H.~Li, and M.~A. Anastasio,
\newblock Approximating the {I}deal {O}bserver and {H}otelling {O}bserver for
  binary signal detection tasks by use of supervised learning methods,
\newblock IEEE Trans. Med. imaging {\bf 38}, 2456--2468 (2019).

\bibitem{chen2012classifier}
W.~Chen, B.~D. Gallas, and W.~A. Yousef,
\newblock Classifier variability: accounting for training and testing,
\newblock Pattern Recognition {\bf 45}, 2661--2671 (2012).

\bibitem{Reiser10}
I.~Reiser and R.~M. Nishikawa,
\newblock Task-based assessment of breast tomosynthesis: effect of acquisition
  parameters and quantum noise,
\newblock Med. Phys. {\bf 37}, 1591--1600 (2010).

\bibitem{schmidt2017spectral}
T.~G. Schmidt, R.~F. Barber, and E.~Y. Sidky,
\newblock A spectral {CT} method to directly estimate basis material maps from
  experimental photon-counting data,
\newblock IEEE Trans. Med. Imaging {\bf 36}, 1808--1819 (2017).

\end{thebibliography}

\end{document}